\documentclass[symmetry,article,accept,moreauthors,pdftex]{Definitions/mdpi}

\newcommand{\aap}{{\it Astron. Astrophys.}}
\newcommand{\aaps}{{\it Astron. Astrophys. Suppl.}}
\newcommand{\araa}{{\it Annu. Rev. Astron. Astrophys.}}
\newcommand{\apj}{{\it Astrophys. J.}}
\newcommand{\apjl}{{\it Astrophys. J. Lett.}}
\newcommand{\apjs}{{\it Astrophys. J. Supp. Ser.}}
\newcommand{\apss}{{\it Astrophys. Space Sci.}}

\newcommand{\mnras}{{\it Mon. Not. R. Astron. Soc.}}
\newcommand{\nat}{{\it Nature}}
\newcommand{\prc}{{\it Phys. Rev. C}}
\newcommand{\prd}{{\it Phys. Rev. D}}
\newcommand{\prl}{{\it Phys. Rev. Lett.}}

\newcommand{\sovast}{{\it Sov. Astronom.}}

\newcommand{\physrep}{{\it Phys. Rep.}}

\firstpage{1} 
\pubvolume{1}
\issuenum{1}
\articlenumber{0}
\pubyear{2022}
\copyrightyear{2021}
\externaleditor{Academic Editor: Kazuharu Bamba} %MDPI: AUTHORS: We have removed highlights 
\datereceived{} 
\dateaccepted{} 
\datepublished{} 
\hreflink{https://doi.org/} % If needed use \linebreak
\pdfoutput=1

\usepackage{todonotes}
\usepackage{caption,subfigure}
\usepackage{bm}
\usepackage{xcolor}
\Title{Magnetic Field Evolution in Neutron Star Crusts: Beyond the Hall Effect}

\TitleCitation{Magnetic Field Evolution in Neutron Star Crusts: Beyond the Hall Effect}

\Author{Konstantinos N. Gourgouliatos  %Please carefully check the accuracy of names and affiliations. Changes will not be possible after proofreading. Authors: This is correct
 $^{1,}$*\orcidA{}, Davide De Grandis $^{2}$\orcidB{} and Andrei Igoshev $^{3}$\orcidC{}}

\AuthorNames{Konstantinos N. Gourgouliatos, Davide De Grandis and Andrei Igoshev }

\AuthorCitation{Gourgouliatos, K.N.; De Grandis, D.; Igoshev, A.}
\address{%
$^{1}$ \quad Department of Physics, University of Patras, 26504 Patras, Greece\\
$^{2}$ \quad Department of Physics and Astronomy, University of Padova, via Marzolo 8, 
I-35131 Padova, Italy; davide.degrandis@phd.unipd.it\\
$^{3}$ \quad School of Mathematics, University of Leeds, Woodhouse, Leeds LS2 9JT, UK; a.igoshev@leeds.ac.uk}

\corres{Correspondence: kngourg@upatras.gr}

\abstract{Neutron stars host the strongest magnetic fields that we know of in the Universe. Their magnetic fields are the main means of generating their radiation, either magnetospheric or through the crust. Moreover, the evolution of the magnetic field has been intimately related to explosive events of magnetars, which host strong magnetic fields, and their persistent thermal emission. The evolution of the magnetic field in the crusts of neutron stars has been described within the framework of the Hall effect and Ohmic dissipation. Yet, this description is limited by the fact that the Maxwell stresses exerted on the crusts of strongly magnetised neutron stars may lead to failure and temperature variations. In the former case, a failed crust does not completely fulfil the necessary conditions for the Hall effect. In the latter, the variations of temperature are strongly related to the magnetic field evolution. Finally, sharp gradients of the star's temperature may activate battery terms and alter the magnetic field structure, especially in weakly magnetised neutron stars. In this review, we discuss the recent progress made on these effects. We argue that these phenomena are likely to provide novel insight into our understanding of neutron stars and their observable properties.}


\begin{document}
%%%%%%%%%%%%%%%%%%%%%%%%%%%%%%%%%%%%%%%%%%

%\begin{paracol}

\section{Introduction}

The magnetic field evolution has been a central topic in the study of the physics of neutron stars (NS). Even before their observational discovery~\cite{Hewish:1968} it had been proposed that NSs host strong magnetic fields which lead to energy emission~\cite{Pacini:1967}. Shortly after the discovery of the first few tens of pulsars, it has been proposed that the distribution of their periods implies that the magnetic field evolves within time-scales of $10^{6}$ years %MDPI: please check if this should be replaced with years, if so, please revise for all. AUTHORS: We have replaced with years.
~\cite{Pacini:1969}. Early studies of the conductivity of the crust of NS due to phonon scattering and impurities~\cite{Ewart:1975} suggested decay times for the magnetic field to be at least of the order of $10^{7}$ years. %MDPI: 
 These results were revised a few years later in more detailed calculations~\cite{Yakovlev:1980}. Despite the high thermal conductivity of the crust, their temperature is neither homogeneous nor isotropic, as~heating is produced in regions where the electric currents are stronger and is primarily conducted along magnetic field lines. The~impact of temperature gradients within the NS has been presented in~\cite{Urpin:1980,1983MNRAS.204.1025B}, where it has been proposed that the magnetic field of NSs can be created due to Biermann battery effects~\cite{1950ZNatA...5...65B}. The~effect of Hall drift in addition to Ohmic decay was discussed in~\cite{Jones:1988}, who proposed that the decay of the magnetic field is mediated by Hall drift. In~the seminal work of~\cite{Goldreich:1992} three main effects have been proposed to operate in NSs affecting the evolution and eventual decay of the magnetic field: Ohmic decay, Hall drift and ambipolar diffusion. The~Ohmic decay is due to the finite conductivity of the crust of the NS, which leads to dissipation of the magnetic field and conversation of energy into thermal power~\cite{1990A&A...229..133H,1998ApJ...503..368M,2006MNRAS.371..477K}. Hall drift is related to the presence of free electrons in the crust which carry the electric current while the ion lattice remains rigid, thus, any Lorentz force remains in equilibrium due to crust elasticity. Finally, ambipolar diffusion is caused by the interaction of the charged particles with the abundant neutrons, leading to the dissipation of the~currents. 

This has motivated several studies aiming to understand the dynamics of magnetic field evolution. Moreover, the~discovery of Anomalous X-ray Pulsars (AXPs) and Soft Gamma-ray Repeaters (SGRs) \cite{1979Natur.282..587M,1981Natur.293..202F, 1987ApJ...320L.111L,1987ApJ...322L..21K,2002Natur.419..142G,2003ApJ...588L..93K}, comprising the magnetar population~\cite{1992ApJ...392L...9D,1998ApJ...506L..61H,2010PNAS..107.7147K,2017ARA&A..55..261K}, has generated a large interest in the overall evolution of the magnetic field in NSs, as~in such sources, it is not the rotational kinetic energy loss rate that dominates, but~rather the magnetic field decay. The~main aims of these studies are to provide a mechanism for the efficient conversion of magnetic field energy into heat and to study the possibility of instabilities that could lead to explosive events. This is because the magnetic field energy reservoir can provide sufficient heat, and~power the luminosity of these objects~\cite{2005A&A...433..275P,2007PhRvL..98g1101P,2008A&A...483..223U,2017PhRvD..96j3012G}. Therefore, their persistent emission can be attributed to their magnetic field. Furthermore, these strong magnetic fields can trigger explosive events which are observed in the forms of bursts, outbursts  and flares~\cite{2018ASSL..457...57G,2021ASSL..461...97E}.

The interplay between the Hall effect and Ohmic decay has been explored in detail using analytical and numerical techniques~\cite{1997A&A...321..685S,2002MNRAS.337..216H,2002PhRvL..88j1103R,2004ApJ...609..999C,2004MNRAS.347.1273H,2007A&A...472..233R,2007A&A...470..303P,2012CoPhC.183.2042V,2012MNRAS.421.2722K,2013MNRAS.434.2480G,2014MNRAS.438.1618G,2015PhRvL.114s1101W,2016PNAS..113.3944G,2016MNRAS.456.4461E,2017JPhCS.932a2048P,2018A&G....59e5.37G,2018ApJ...852...21G,2019LRCA....5....3P,2020ApJ...901...18B,2020MNRAS.495.1692G,2021CoPhC.26508001V}. Apart from the isotropic conductivity of the lattice, pasta phases such as rods, slabs, tubes and bubbles, appearing at the base of the crust, where the nuclear forces and charge screening play an important role lead to enhancements of the resistivity and changes to its overall properties~\cite{2013NatPh...9..431P,2021PhRvC.103e5812X}.
These models have successfully addressed observational properties of NSs and they have revealed, moreover, rich effects in terms of magnetohydrodynamical evolution, arising from the non-linear nature of the equations. These effects include instabilities and turbulent cascades and are of broader applications within the realm of magnetohydrodynamics~\cite{1999PhPl....6..751B,2004A&A...420..631R,2009PhPl...16d2307W,2009A&A...508L..39W,2009ApJ...701..236C,2009A&A...496..207P,2010JPlPh..76..117W,2010A&A...513L..12P,2014ApJ...796...94M,2014PhPl...21e2110W,2015MNRAS.453L..93G,2016MNRAS.463.3381G,2017AstL...43..624K,2019PhRvR...1c2049G,2019AN....340..475K,2021JPlPh..87d9004K}. 

The assumption of Hall and Ohmic evolution as formulated in~\cite{Goldreich:1992} is valid for NSs whose magnetic field is moderately strong. NSs with very strong magnetic fields are susceptible to crust failure, as~deformations beyond the crust elastic limit lead to permanent deformation~\cite{1991ApJ...382..587R,2009PhRvL.102s1102H,2010MNRAS.407L..54C,2011MNRAS.416...22B,2012MNRAS.426.2404H,2012CoPP...52..122C,2013PhRvD..88d4004J,2018MNRAS.480.5511B,2020MNRAS.491.1064G}. Thus, the~assumption of a rigid crust that can absorb the stresses induced by the magnetic field and maintain its equilibrium does not hold any more.  The~failure of the crust may be due to magnetic field strength and be associated to magnetar activity~\cite{2011ApJ...741..123P,2011ApJ...727L..51P, 2020ApJ...902L..32D, 2020ApJ...903...40D} or even due to binary interaction~\cite{2012PhRvL.108a1102T,2013ApJ...777..103T,2021arXiv211103686N}. Therefore, in the regime of strong magnetic fields, the~crust of the NS can no longer be treated as a rigid background and the description due to the Hall effect should be complimented to account for these~effects. 

The evolution of the magnetic field in NSs with weaker fields, on~the contrary, is not dominated by the Hall effect, as~the associated timescale is much longer.  In~this regime other causes of magnetic field evolution may be critical such as Biermann Battery. Independently of the magnetic field strength, thermal anisotropies can lead to interplay between magnetic and thermal terms in the evolution and drive magnetic field evolution. The~electric conductivity increases when NS cools down and phonons do not contribute to the resistivity anymore~\cite{Igoshev2015AN,Igoshev2014MNRAS}. 
Moreover, deeper in the crust, the~effects of ambipolar diffusion become critical where the electric currents interacts, yet as this effect is mostly relevant to the physics of the NS core, it will not be discussed in detail in this~review.

The aim of this review is to present the conditions under which the Hall effect is no longer dominant in the crust of NSs, and~also the evolution regimes dominating beyond the Hall effect. In~particular, we will be focusing on the effects of plastic evolution, magnetothermal evolution and battery~effects. 

In Section~\ref{sec:Hall} we present the basic theory of the Hall-Ohmic evolution and the region of applicability. In~Section~\ref{sec:Plastic} we present the effect of Plastic evolution. In~Section~\ref{s:magnetothermal} we present the effects related to magneto-thermal evolution and, in~particular, the~thermal battery effects are addressed in Section~\ref{sec:battery}. We conclude in Section~\ref{sec:conclusions}.

\section{Hall-Ohmic~Evolution}
\label{sec:Hall}

The crust of a NS comprises an ion lattice and free electrons while neutrons become abundant at densities deeper than the neutron drip point marking the transition from the outer to the inner crust~\cite{2008LRR....11...10C}. The~electrons have the freedom to move within the crust and are the carriers of the electric current. On~the contrary, ions stay at fixed locations, as~long as the crust remains rigid. This allows a single fluid MHD-description, with~a direct relation between the electric current and the electron fluid velocity. The~rigidity of the ion lattice is sustained provided that the crust remains within its elastic limit in terms of the deformations that are due to the stresses acting upon the crust. Therefore, the~momentum equation is identically satisfied and the evolution of the system is described through the following Hall-Ohmic induction equation:
\begin{eqnarray}
\partial_{t} {\vec B} = -{\vec \nabla}\times \left[\frac{c}{4 \pi e n_e}\left({\vec \nabla} \times {\vec B}\right)\times {\vec B}  +\frac{c^2}{4 \pi \sigma}{\vec \nabla} \times {\vec B}\right]\,,\
\label{HALL_EQ}
\end{eqnarray}
where $\vec{B}$ is the magnetic field, $e$  is the electron charge, $n_{e}$ is the electron number density, $\sigma$ is the electric conductivity and $c$ is the speed of light. The~first term on the right hand side is the Hall term, describing the advection of the magnetic field by the electron fluid, while the second one is the Ohmic dissipation term. While the Hall term is conservative, it facilitates and accelerates magnetic field decay, as~it leads to the formation of small-scale structures, which dissipate faster due to Ohmic decay. The~above equation is a non-linear partial differential equation which bears some resemblance to the vorticity equation from fluid dynamics. This has motivated several works studying the effects of electron-MHD turbulence~\cite{1999PhPl....6..751B,2009ApJ...701..236C,2009PhPl...16d2307W,2009A&A...508L..39W}. A~remarkable effect that differentiates the electron-MHD evolution from the typical fluid turbulence is the fact that while a turbulent spectrum develops, the~temporal evolution freezes-out and the field maintains its structure, attaining a Hall-attractor state~\cite{2014PhRvL.112q1101G}. 

The implications of Hall evolution in NSs crusts have been rather profound. First, it has been shown that the magnetic field evolves faster once its strength is higher. Thus, magnetars, not only have a larger reservoir of magnetic energy, but~they convert it faster into heat. This effect can be directly related to the steep dependence of the X-ray bolometric power on the dipole magnetic field strength~\cite{2009A&A...496..207P,2013MNRAS.434..123V,2014ApJS..212....6O}. Second, the~magnetic field evolution is rapid in young NSs and results in explosive events. The~evolution later saturates, leading to the dichotomy between magnetars and older NSs shining in X-rays~\cite{2011ApJ...741..123P,2014PhRvL.112q1101G,2018ASSL..457...57G}. Third, the~magnetic field evolution leaves its imprint on the timing evolution of NSs, either in the form of enhanced noise, or~through their braking indices~\cite{2012A&A...547A...9P,2013ApJ...773L..17T,2015MNRAS.446.1121G,2016MNRAS.456...55G,2017ApJ...849...19G,2019ApJ...880..123X,2020EPJC...80..411C}. Fourth, the~evolution of the magnetic field leads to the spontaneous formations of spots, arcades and multipolar structures~\cite{2013MNRAS.435.3262G,2014MNRAS.444.3198G,2015ApJ...807L..20B,2018ApJ...852...21G,2019PhRvR...1c2049G} which have been related to magnetars and have been directly observed in several NSs, either directly or as spectral features~\cite{2003Natur.423..725B,2011MNRAS.418.2773G,2013Natur.500..312T,2019A&A...626A..39P,2019MNRAS.489.4589A}. Fifth, apart from the obvious connection to strongly magnetised NSs, the~Hall effect has been related to Central Compact Objects~\cite{2005ApJ...627..390G,2007ApJ...664L..35G,2009ApJ...695L..35G,2010ApJ...709..436H,2013ApJ...765...58G,2014ApJ...792L..36B,2017JPhCS.932a2006D}, in~the scenario of the re-emergence of the buried magnetic field. In~this context, the~magnetic field has been buried following the formation of the NS, and~while it provides heat so that the star shines in X-rays, its dipole strength is relatively weak~\cite{2012MNRAS.425.2487V,2011MNRAS.414.2567H,2016MNRAS.462.3689I,2020MNRAS.495.1692G,Igoshev2021ApJ}. Finally, the~role of the Hall effect has been central in the study of the entire NS population. The~magnetic field evolution has been a major ingredient of the grand unification of NSs~\cite{2013MNRAS.434..123V,2014MNRAS.443.1891G,2015MNRAS.454..615G}.

Therefore, the~detailed study of the magnetic field evolution due to the Hall effect has been proven fruitful and successful. It has not only provided a rich range of effects from a theoretical perspective, but~it has also led to direct applications to observable properties of NSs, not only in the range of the strongly magnetised ones but across the entire NS~population.

\section{Plastic~Evolution}
\label{sec:Plastic}

The plethora of explosive events accompanying strongly magnetised NSs is strong evidence of the rapid and drastic changes happening in NSs. The~various scaling relations between the magnetic field strength and the intensity of the X-ray bolometric luminosity and the power emitted in outbursts~\cite{2018MNRAS.474..961C} is evident that these phenomena are driven by the magnetic field. The~energy emitted in explosive events~\cite{1999Natur.397...41H,2005Natur.434.1098H,2005Natur.434.1107P} can be stored in the form of twisted field lines in the magnetosphere and released through explosive reconnection~\cite{1986ApJ...307..205L,1991ApJ...375L..61A,1997PhR...283..185C,2003MNRAS.346..540L,2006MNRAS.367.1594L,2007ApJ...657..967B,2007MNRAS.374..415K,2008MNRAS.385..875G,2008MNRAS.391..268G,2009ApJ...703.1044B,2010GApFD.104..431G,2012MNRAS.423.1416P,2013ApJ...774...92P,2014ApJ...784..168H,2018MNRAS.474..625A,2019MNRAS.490.4858M}. Alternatively, the~energy can be stored within the crust directly in the form of magnetic field energy or elastic energy due to the deformation caused by the magnetic field~\cite{1995MNRAS.275..255T,2011ApJ...727L..51P,2012MNRAS.427.1574L,2014ApJ...794L..24B,2015MNRAS.449.2047L,2016ApJ...833..189L,2019MNRAS.488.5887S,2021MNRAS.506.3936K}. An~explosive event in the latter case can occur if the crust becomes deformed beyond its elastic limit. The~details of the evolution of the magnetic field once the crust has failed depend strongly on the properties of the crust and the exact process following the crust failure. We can obtain a generic view of the evolution of a failed crust through the Navier-Stokes equation~\cite{2016ApJ...824L..21L}:
\begin{eqnarray}
\rho \dot{\vec v} +\rho\left({\vec v}\cdot {\vec \nabla}\right){\vec v}= -{\vec \nabla} P -\rho {\vec \nabla} \Phi +\nu \nabla^2 {\vec v}+\frac{1}{4\pi}\left({\vec \nabla} \times {\vec B}\right)\times {\vec B} +{\vec \nabla} \cdot \hat{ \tau}_{\rm el}\,,
\end{eqnarray}
where $\rho$ is the density, $\vec{v}$ is the plastic flow velocity, $P$ is the pressure, $\Phi$ is the gravitational potential, $\nu$ is the viscosity and $\hat{\tau}_{\rm el}$ is the elastic stress tensor. In~the above equation the gravitational potential gradient and the radial pressure gradient are by far the strongest terms, and~considering a stably stratified crust, they would not lead to radial velocities once perturbed by the other terms. Possibly, a~strong magnetic field and rapid rotation may lead to deformation from spherical symmetry. Moreover, the~plastic flow velocity is anticipated to be slow, thus the quadratic term and the acceleration terms can be safely neglected. As~the crust is incompressible,  the~continuity equation reduces to the following expression for the velocity:
\begin{eqnarray}
\vec{\nabla} \cdot \vec{v}=0\,.
\end{eqnarray}

The momentum equation can be further simplified if one considers the fact that the crust freezes in a stressed state determined by the initial magnetic field. As~the magnetic field evolves, the~amount of stress exerted onto the crust changes, and~it could lead to deformations exceeding the elastic limit. In~this case, the~failed crust will adopt a plastic flow caused by the excess stress compared to the initial state which we set a reference state. An~important aspect is that the diagonal terms of the stress tensor (volumetric), are related to compression of the crust, whereas the off-diagonal ones (deviatoric) are relevant to shearing terms. Thus, the~relevant part of the Maxwell stress tensor that leads to crust failure is the following:
\begin{eqnarray}
\tilde{\cal M}_{ij}=\frac{1}{4\pi }\left(B_i B_j -\frac{1}{3} B^2\delta_{ij}\right)\,,
\end{eqnarray}
where $B_i$ is the $i$-th component of the magnetic field, and~$\delta_{ij}$ is the Kronecker delta. As~the plastic flow is rather constrained, and~the evolutionary time-scale is long compared to the dynamical time-scale, the~momentum equation is satisfied for an appropriate velocity profile so that the viscous term equates the divergence of the difference between the current state and the reference state of the field~\cite{2019MNRAS.486.4130L}:
\begin{eqnarray}
4\pi \nu \nabla^2 \vec{ v}_{pl}=-4\pi \vec{\nabla}\cdot \left({\hat{\tilde{\cal M\hphantom{,}}}}-{\hat{\tilde{\cal M\hphantom{,}}}}_0\right)\,. 
\label{EQ_STRESS}
\end{eqnarray}

Consequently, the~crust is no longer in a fully rigid state, for~the Hall Equation~(\ref{HALL_EQ}) to hold. Nevertheless, a~modification of the induction equation describes the magnetic \mbox{field evolution:}
\begin{eqnarray}
\partial_{t} {\vec B} = -{\vec \nabla}\times \left[\left( \frac{c}{4 \pi e n_e}{\vec \nabla} \times {\vec B}-\vec{v}_{pl}\right)\times {\vec B}  +\frac{c^2}{4 \pi \sigma}{\vec \nabla} \times {\vec B}\right]\,,
\label{PLASTIC_EQ}
\end{eqnarray}
where the plastic flow velocity $\vec{v}_{pl}$ is determined by Equation~(\ref{EQ_STRESS}). 

A further ingredient needed for the solution of the problem is the conditions under which the crust fails. The~microphysics of the crust allows the determination of the critical value of $\tau_{el}$, above~which the crust is no longer in the elastic regime. This is implemented through the von Mises criterion:
\begin{eqnarray}
\tau_{el}\leq \sqrt{\frac{1}{2}\tilde{\tau}_{ij}\tilde{\tau}_{ij}}\,,
\end{eqnarray}
where $\tilde{\tau}_{ij}$ is the deviatoric part of the stress tensor, which for the Maxwell stresses takes the following form:
\begin{eqnarray}
\tau_{el}\leq \frac{1}{4\pi} \sqrt{\frac{1}{3}B_0^4+\frac{1}{3}B^4+\frac{1}{3}B_0^2B^2-\left(\vec{B}\cdot \vec{B}_0\right)^2}\,,
\label{tau_el}
\end{eqnarray}
where $\vec{B}_0$ is the reference magnetic field corresponding to the freezing state of the crust and $\vec{B}$ is its current~value. 

The value $\tau_{el}$ can be determined by the microphysics of the crust, therefore, from \mbox{Equation~(\ref{tau_el})} one can find the regions where the crust fails. The~failure is not necessarily concentrated only in the region determined by the von Mises criterion, but~it can affect a larger fraction of the star~\cite{2016ApJ...833..189L,2017ApJ...841...54T}. Nevertheless, it is unlikely that a failure can affect the entire crust and it will be more likely a local phenomenon, with~the probable exception of giant flares. Following that, the~magnetic field evolution Equation~(\ref{PLASTIC_EQ}) can be integrated and provide the value of the magnetic field throughout the~evolution. 

\begin{figure}[H]

    \includegraphics[width=0.99\linewidth]{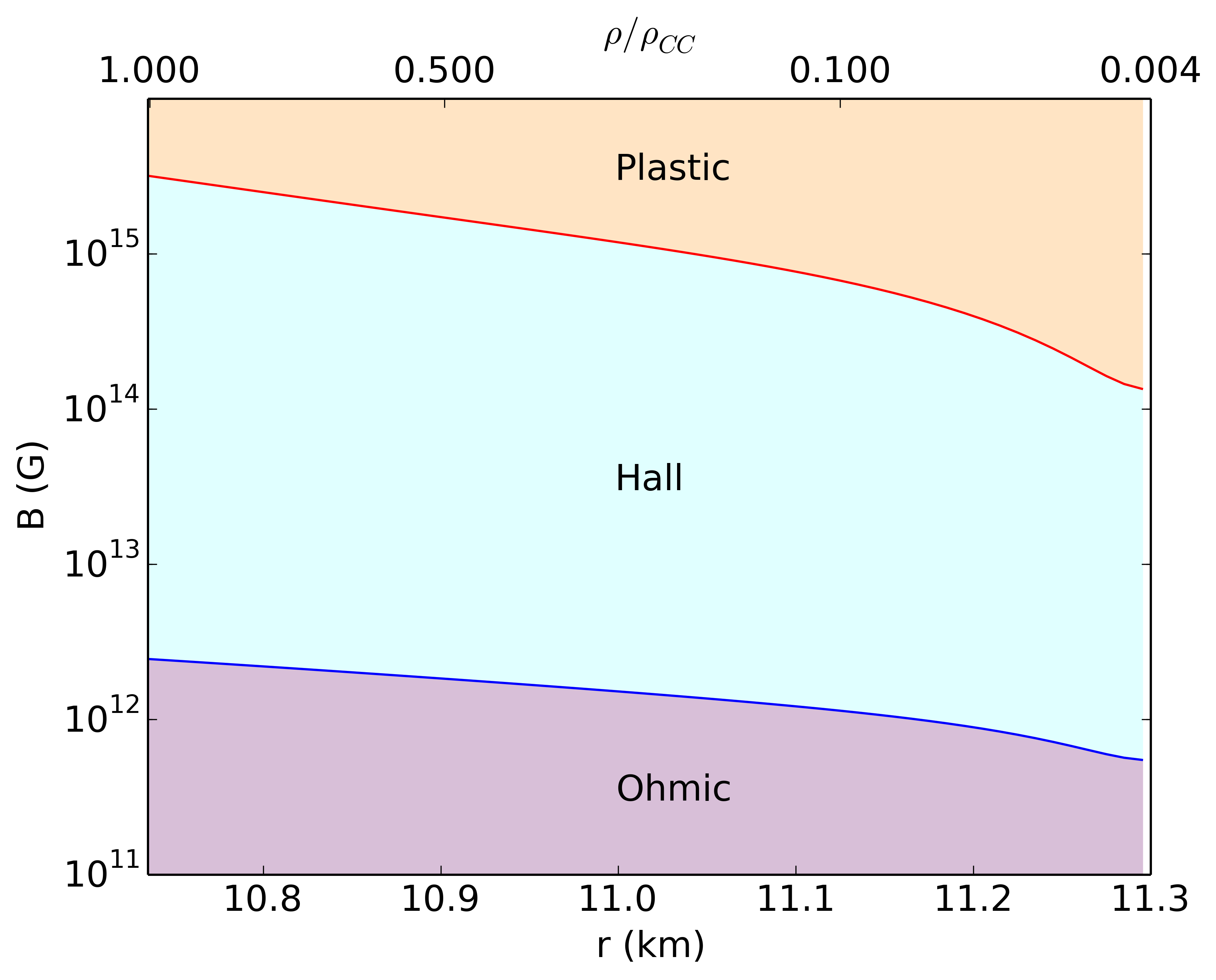}
    \caption{Dominance of Plastic flow, %Please make sure that permission has been obtained and there is no copyright issue. AUTHORS: We have confirmed that the there is no copyright issue. We have confirmed that the there is no copyright issue. The copyright is held by the authors, (corresponding author: K.N. Gourgouliatos), who is also a co-author of the current paper. 
 Hall evolution and Ohmic decay as a function of the magnetic field strength and the radial distance from the centre. The~left boundary starts at the crust-core interface and the right at the neutron drip point. Figure adapted from~\cite{2021MNRAS.506.3578G}.}
    \label{fig:parameters}
\end{figure}

The value of $\tau_{el}$ as a function of the microphysical parameters is the following \cite{2010MNRAS.407L..54C}: % AUTHORS: We have added a citation to this paper as it is relevant to the above equation. 
\begin{eqnarray}
\tau_{el}=\left(0.0195-\frac{1.27}{\Gamma-71}\right)\frac{Z^2 e^2 n_{\rm I}}{a_{\rm I}}\,,
\end{eqnarray}
where $Z$ is the atomic number, $\Gamma$ the Coulomb parameter, $n_{\rm I}$ the ion number density and~$a_{\rm I}$ the ion sphere radius. Using the above relation and the ratio of the Hall to the Ohmic timescales, it is possible to find the combinations of magnetic field strength and density, or~equivalently radius, where the crust fails. Moreover, comparing the Hall and  Ohmic timescales one can find the region in the parameter space where the field evolution is driven by the Hall effect and Ohmic decay. These results are summarized in Figure~\ref{fig:parameters}. It is evident that this effect is crucial near the surface of the star, as~the value of $\tau_{el}$ drops drastically there, yet the strength of the magnetic is not substantially weaker~\cite{2020MNRAS.494.3790K}.

\begin{figure}[H]
\begin{tabular}{ccc}
\includegraphics[height=0.4\linewidth]{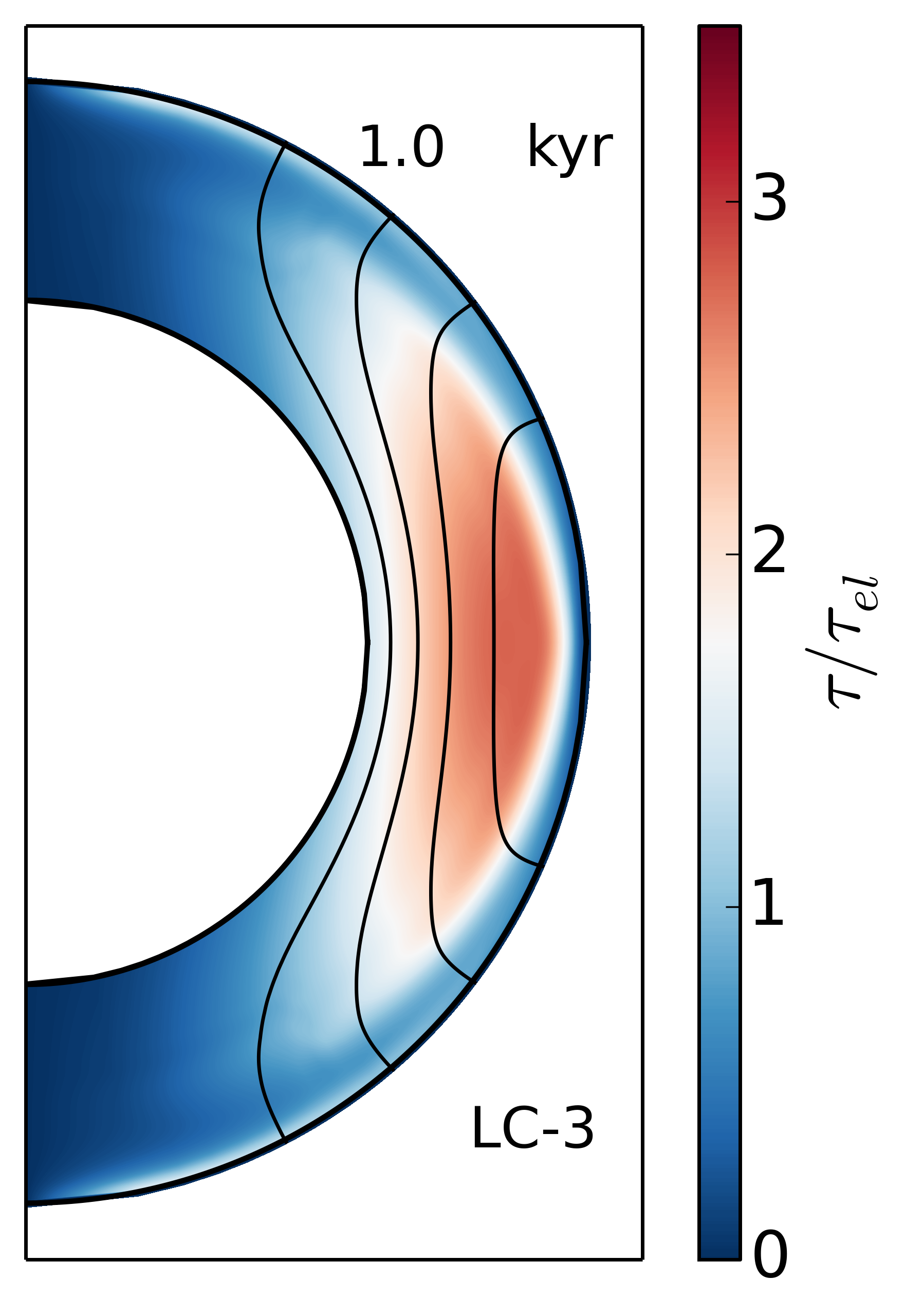}
&\includegraphics[height=0.4\linewidth]{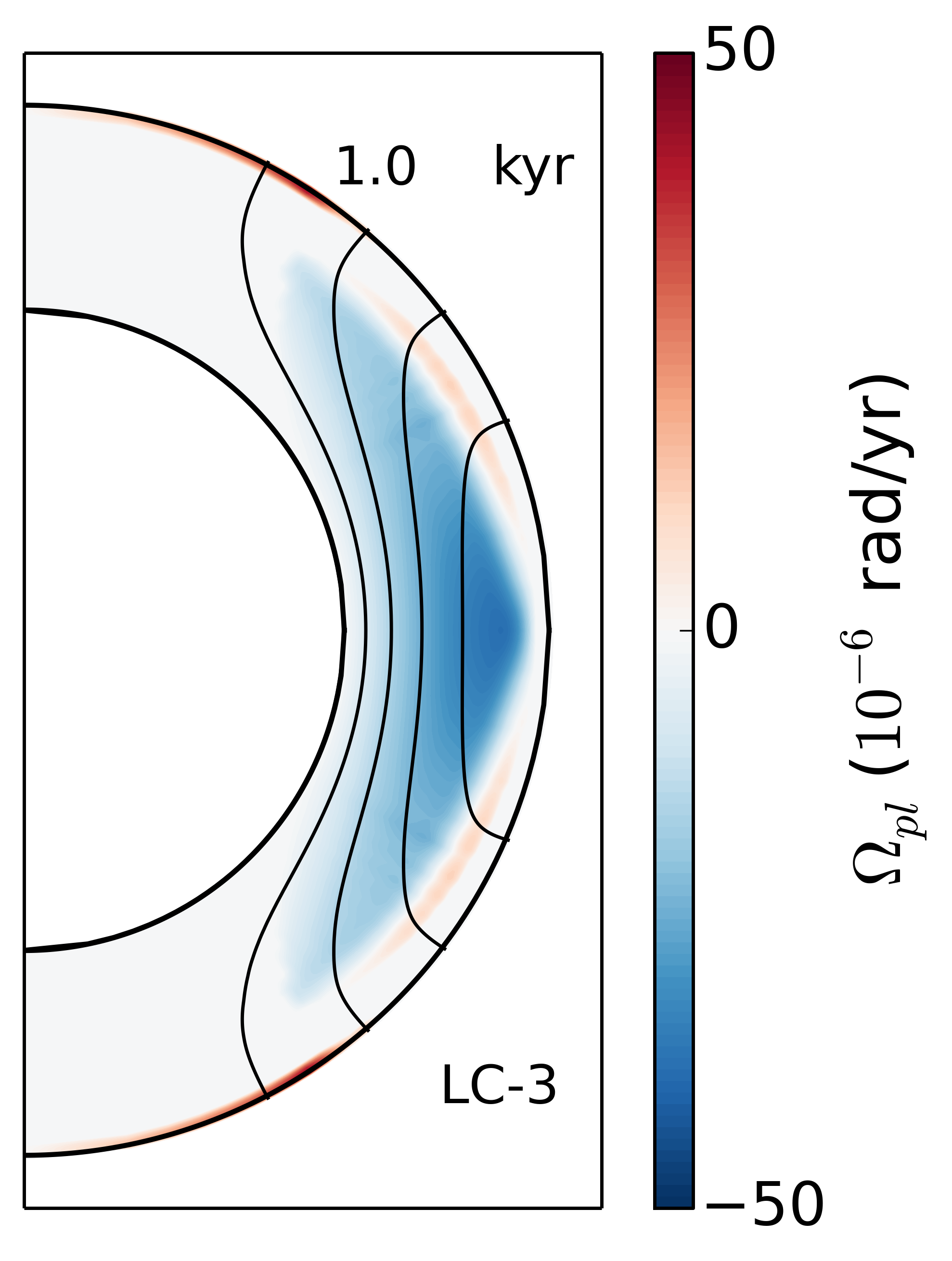}
&\includegraphics[height=0.4\linewidth]{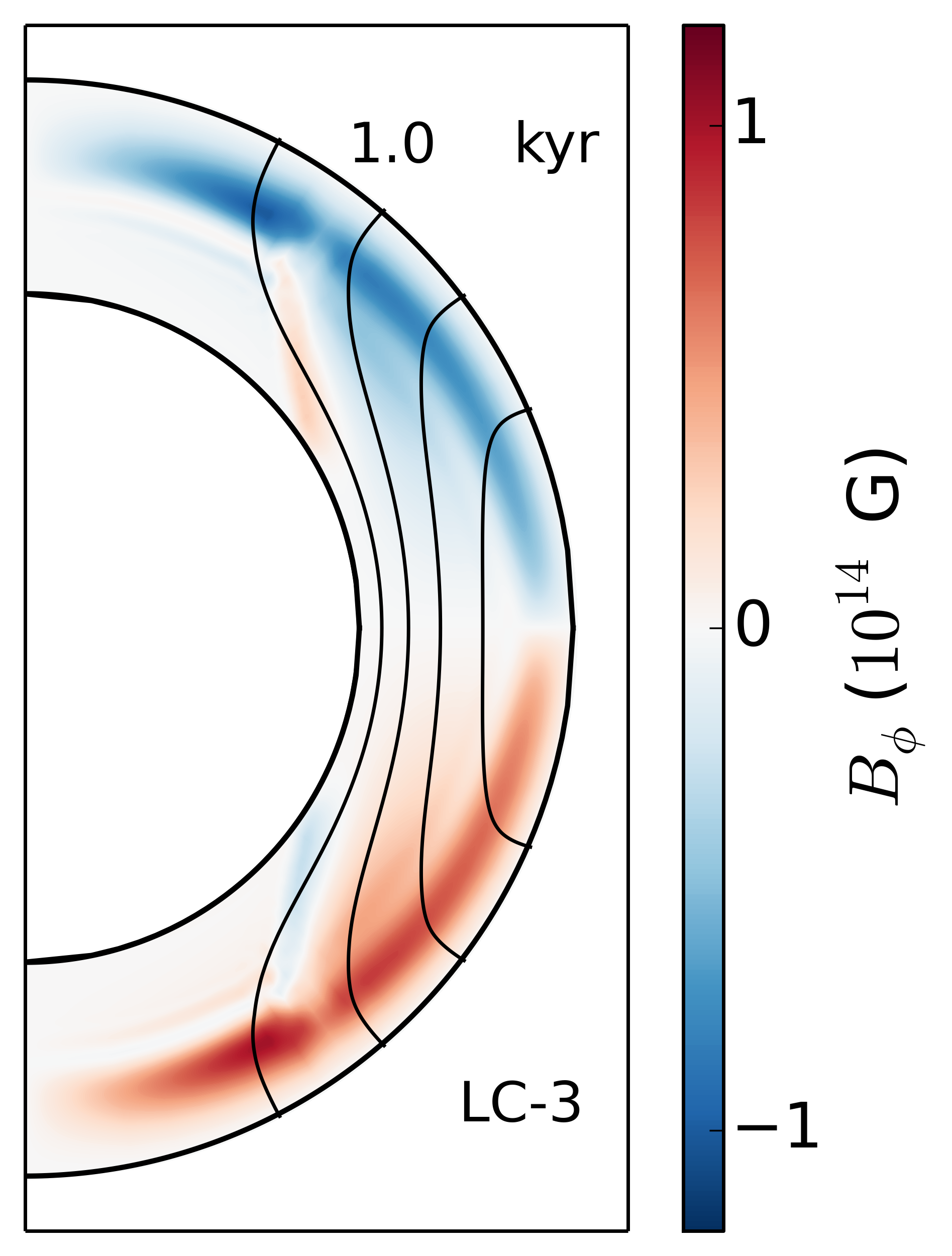}\\
({\bf a})&({\bf b})&({\bf c})\\
\end{tabular}
    \caption{Evolution with plastic flow of the magnetic field at age of 1 kyr.  In~these models a local failure is modelled. The~black lines are the poloidal field lines. The~crust has been expanded for for clarity, in~reality it corresponds to $0.05$ of the stellar radius. In~panel (\textbf{a}) the stresses that have been developed are shown, in~panel (\textbf{b}) the plastic flow velocity and in panel (\textbf{c}) the toroidal field in the crust. Figure adapted from~\cite{2021MNRAS.506.3578G}.}
    \label{fig:plastic_flow}
\end{figure}

Thus, these suffice for the solution of the magnetic field evolution equation. The~results of these solutions have been presented recently describing the evolution of the magnetic field in stressed crusts flowing plastically~\cite{2016ApJ...824L..21L,2019MNRAS.486.4130L,2020MNRAS.494.3790K,2021MNRAS.506.3578G,2020MNRAS.494.3790K,2021MNRAS.502.2097K}. An~example is shown in Figure~\ref{fig:plastic_flow}. The~main conclusions are the following:
\begin{enumerate}
    \item For strong magnentic fields, $B>10^{15}$ G, %MDPI: space added, please confirm. AUTHORS. We confirm, a space is needed
 plastic flow occurs everywhere in the crust and large parts of the crust fail. 
    \item For magnetic fields of strength $B<10^{14}$ G, plastic flow essentially shuts down. 
    \item For magnetic fields between $10^{14}{\rm G }<B<10^{15}$ G, failure occurs sporadically in various crust regions. 
    \item In general, plastic flow opposes the evolution generated by the Hall effect. Nevertheless, it does not completely annul the magnetic field evolution generated by the Hall effect, even under strong magnetic fields. 
    \item The evolution is sensitive to the parameter of the plastic flow viscosity. A~plastic viscosity of $10^{38}$  g cm$^{-1}$~s$^{-1}$ leads to velocities of $10-100$  cm/yr. 
    \item While plastic flow is a dissipative process, it does not expedite drastically the decay of the magnetic field as it opposes the creation of small-scale fields that would dissipate faster due to the Ohmic effect. 
    \item The results are sensitive to the type of failure, whether it affects a small part of the crust (local failure) or a more extended one (global failure). 
    \item Numerical simulations and analytical solutions have explored so far simple geometries: cartesian plane-parallel and axisymmetric ones. Thus, the~full 3D evolution due to plastic flow may be complex than the results found so far. 
\end{enumerate}

\vspace{-12pt}

\vspace{-6pt}

\section{Magneto-Thermal~Evolution}

\label{s:magnetothermal}

Both electric current and thermal flux are transported by electrons in the NS crust. In~a strong magnetic field, electrons can freely move along the magnetic field lines but their motion in the direction orthogonal to magnetic field lines is strongly suppressed. Thus, heat is easily transported only along the magnetic field lines. Therefore, two distant regions connected by field lines have the same temperature. On~the contrary, two geometrically close regions separated by magnetic fields could have a large temperature~difference.

The equation for heat transfer is written as follows
\begin{equation}
C_v \frac{\partial T}{\partial t} = \vec \nabla \cdot (\hat k \cdot \vec \nabla T) + \frac{|\vec \nabla \times \vec B|^2 c^2}{16\pi^2 \sigma} + \left(\frac{c}{4\pi e}\right) T \vec \nabla S_e \cdot (\vec \nabla \times \vec B)    \label{eq:heat}
\end{equation}
where $T$ is the temperature, $C_v$ is the heat capacity, $S_e$ is the electron entropy and $\hat k$ is the thermal conductivity tensor, which depends on the magnetic field strength and direction. The~first term on the right-hand-side describes anisotropic heat transfer; the second term corresponds to heat released due to Ohmic losses in the NS crust and the last term is the thermopower one. 

\begin{figure}[H]
    \includegraphics[width=0.85\linewidth]{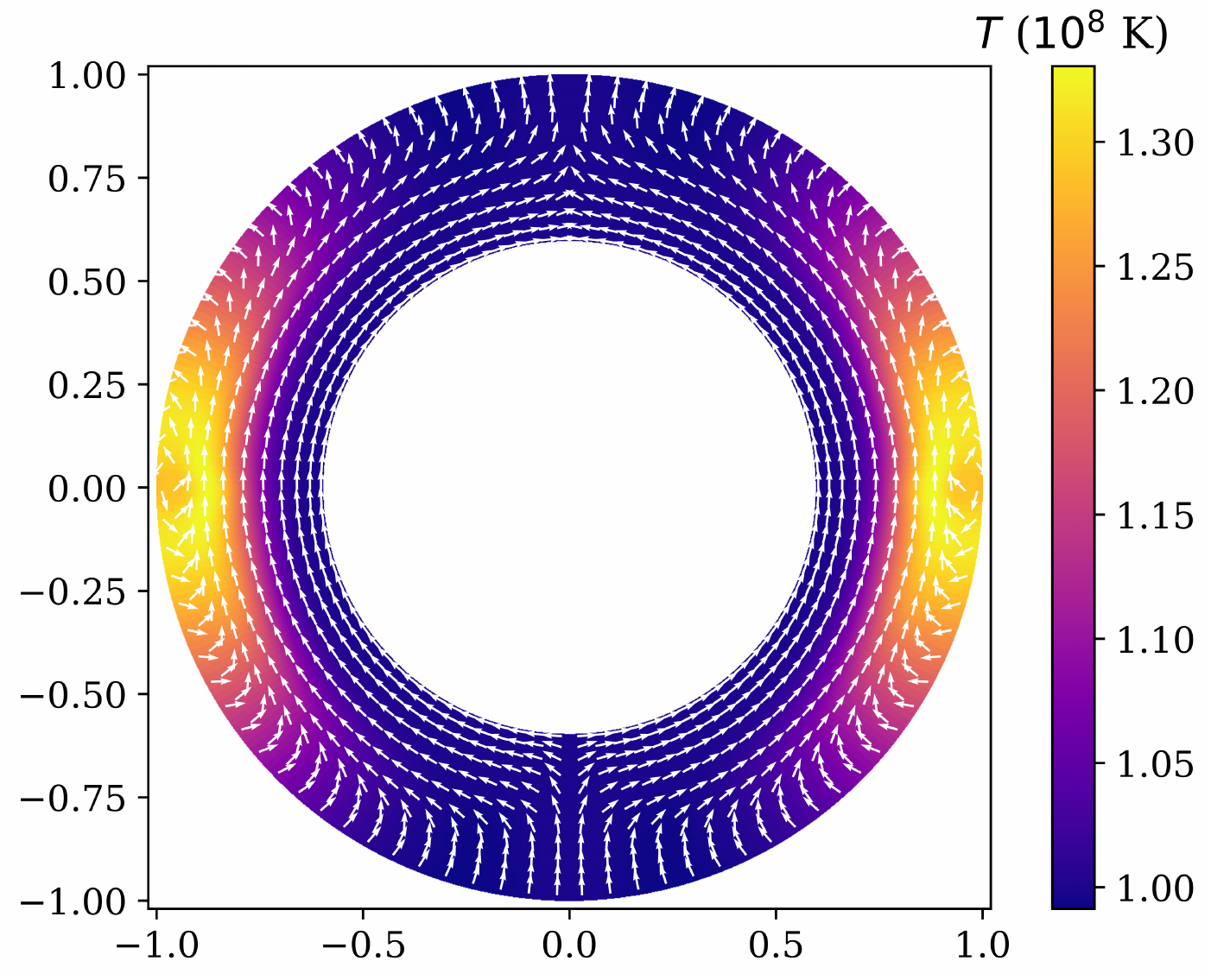}
    \caption{Magneto-thermal structure in vertical slice of NS at age of $9$\,kyr. The~initial magnetic field configuration consists of only a dipolar poloidal magnetic field. The~orientation of the magnetic field is shown only in every 12th radial point and every 4th angular point. White arrows show the direction of $B_r, B_\theta$ magnetic field. The~thickness of the NS crust is scaled up to show details of the magneto-thermal structure. The~equatorial region is thermally isolated from the colder~core.}
    \label{f:t_B}
\end{figure}

\begin{figure}[H]
    \centering
    \includegraphics[width=0.97\linewidth]{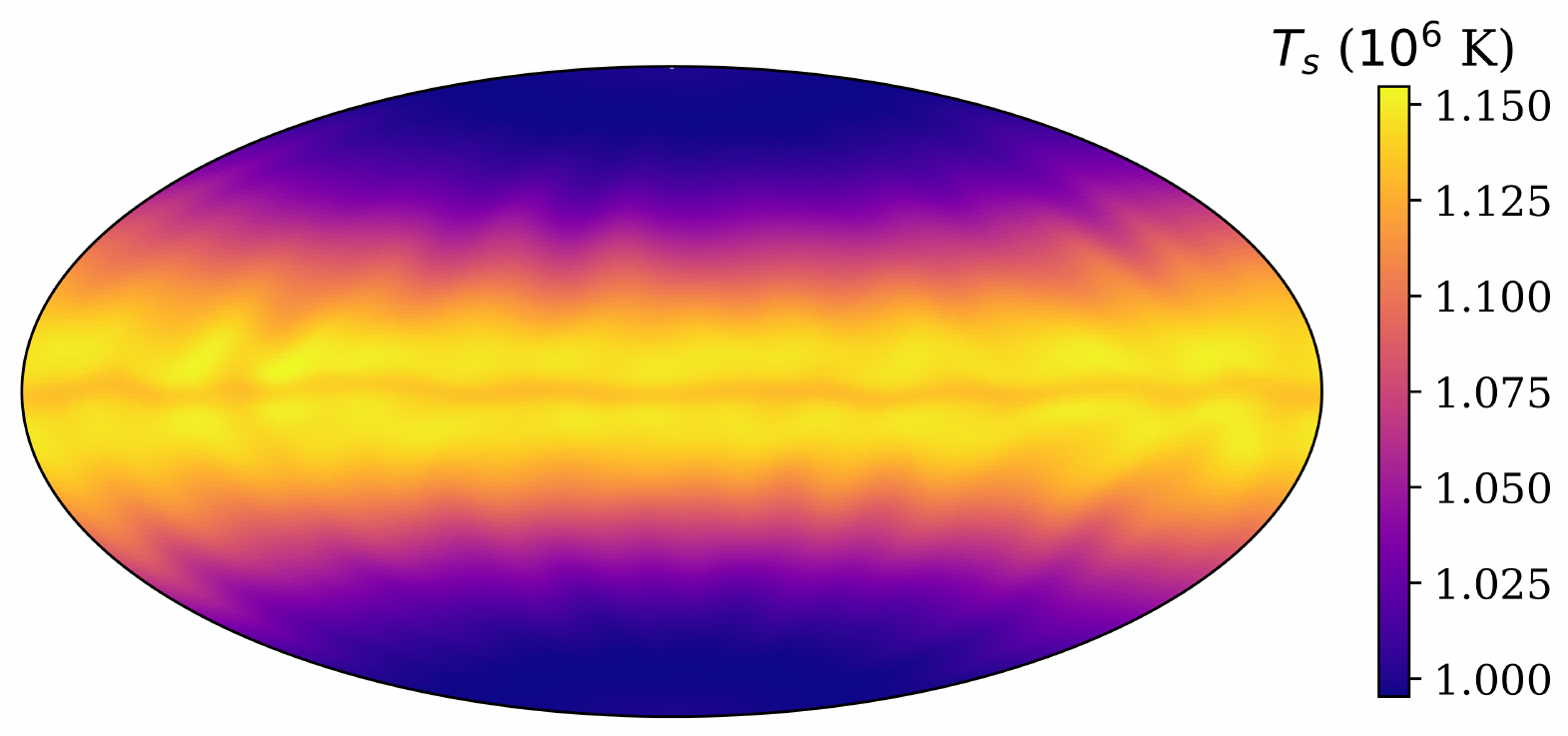}
    \caption{Surface temperature distribution for NS at age $9$\,kyr. The~initial magnetic field configuration consists of only dipolar poloidal magnetic~field.}
    \label{f:surf_T}
\end{figure}

Even in the case of relatively simple magnetic field configuration such as the poloidal dipolar magnetic field, certain regions become thermally isolated from the NS core. In \mbox{Figure~\ref{f:t_B}} we show the temperature and magnetic field configuration inside the NS crust computed using the PARODY code \citep{Dormy1998,2015PhRvL.114s1101W,2016PNAS..113.3944G,Igoshev2021NatAs,2020ApJ...903...40D,Igoshev2021ApJ, 2021ApJ...914..118D}. In~Figure~\ref{f:surf_T} we show the corresponding surface temperature distribution. We obtain these surface temperature distributions using the following simplified relation \citep{GudmundssonPethick1983ApJ} between the surface temperature $T_s$ and the temperature at the bottom of the envelope $T_b$
\begin{equation}
\frac{T_b}{10^8\; \mathrm{K}} = \left(\frac{T_s}{10^6 \; \mathrm{K}}\right)^2    
\end{equation}

\vspace{-14pt}

In this configuration the equatorial regions become thermally decoupled from the NS core. These regions could be heated due to the ohmic decay of the magnetic field creating a hot belt around the NSs. At~the same time, these regions could also become colder than the polar regions due to the more efficient neutrino cooling and smaller heat capacity of the crust \citep{2013MNRAS.434..123V}. Therefore, the~surface thermal distribution is strongly affected by magnetic field configuration in the NS crust. The~surface temperature distribution is an important ingredient to model the quiescent emission of magnetars and eventually to probe their magnetic fields and emission properties; see for example a recent review~\cite{Igoshev2021Univ}. 

\begin{figure}[H]
    \centering
    \includegraphics[width=0.97\linewidth]{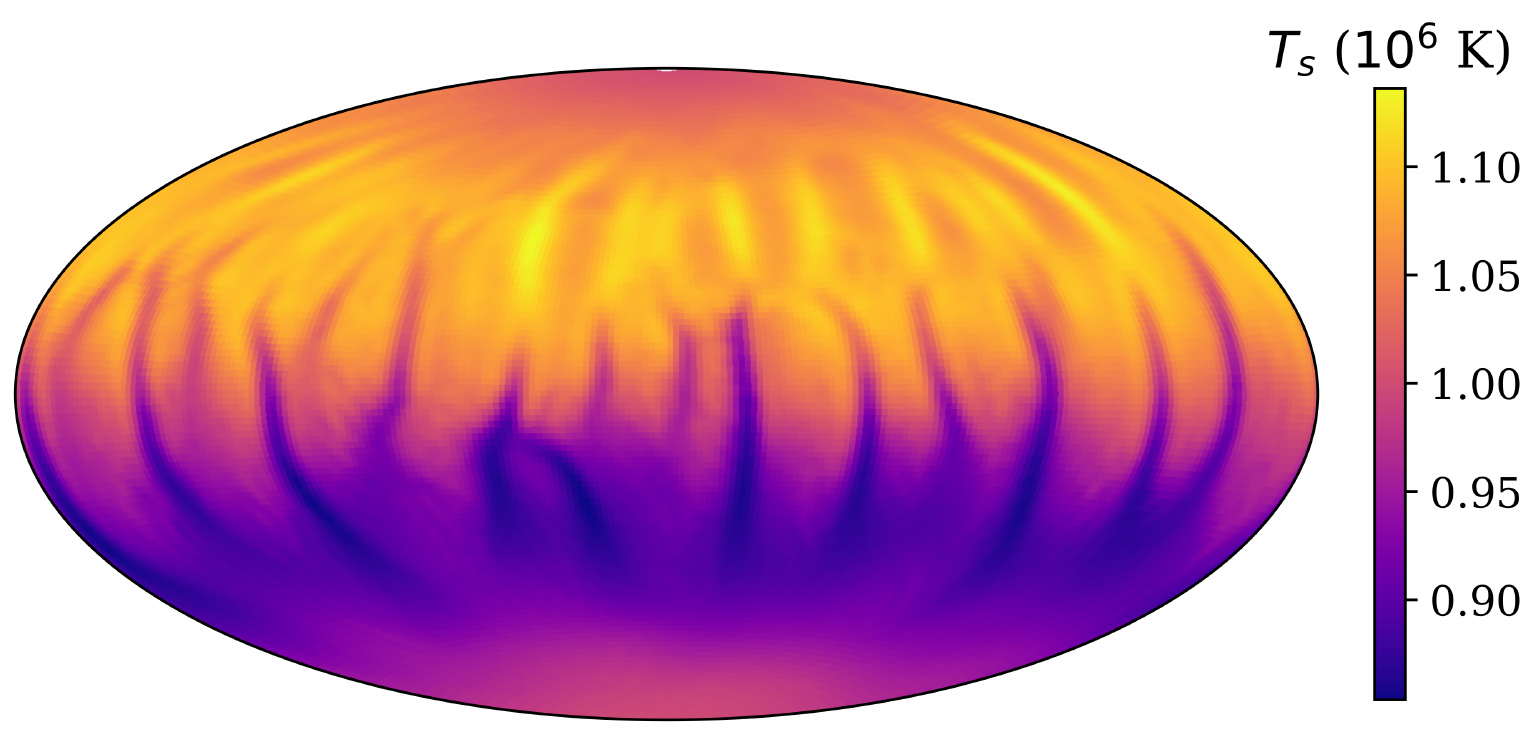}
    \caption{Surface temperature distribution for a NS at age of $40$~kyr. The~initial magnetic field configuration consists of a dipolar poloidal magnetic field and a toroidal magnetic field which contains $99\%$ of the total magnetic~energy.}
    \label{f:t99}
\end{figure}

Moreover, the~Hall evolution couples poloidal and toroidal magnetic fields. For~example, in~a case where the toroidal magnetic component is much stronger than the poloidal one, we observe the toroidal field instability \citep{2019PhRvR...1c2049G} that leads to a significant increase of small scale poloidal magnetic fields in the NS crust. Thus, this instability leaves footprints at surface temperature distribution. Namely, it leads to formation of long hot or cold filaments extending from the northern hemisphere to southern hemisphere; see example in Figure~\ref{f:t99}.  

\vspace{-6pt}

It is quite uncertain at the moment what the exact initial conditions are for the internal magnetic fields of NSs. The~initial magnetic fields could be large-scale (dipole or quadropole) or, instead, it could consists of a superposition of modes with length-scales comparable to the NS crust depth. Such configurations are predicted e.g.,\ by models of stochastic dynamos \citep{ThompsonMurray2001ApJ}. On~the other hand, it is thought that large scale magnetic fields are generated if the NS initially spins fast, with~periods comparable to a few msec. In~reality, initial periods of radio pulsars with strong fields $>10^{12}$\,G range from msec values to $\approx 0.5$\,s~\cite{IgoshevPopov2013MNRAS,PopovTurolla2012Ap}.  

In general, the~large scale magnetic field creates the surface temperature distribution with hot and cold regions of relatively large size (typically comparable to half of the total surface). Typically, the~surface thermal distributions produced in these cases have a high degree of axial symmetry. Only specific initial conditions, such as an addition of toroidal magnetic field inclined with respect to the poloidal component, could cause the appearance of relatively small hot regions~\cite{Igoshev2021NatAs}. Such a configuration could explain a large pulsed fraction (up to 50\%) for thermal X-ray emission which is observed in some \mbox{quiescent magnetars~\cite{Igoshev2021NatAs}.} 

In the case of stochastic dynamos, when the whole magnetic field consists mostly of small-scale fields, the~surface temperature distribution becomes patchy with small hot and cold spots with typical size $\sim$1\,km. At~any orientation of the NS, a~distant observer sees multiple hot and cold regions simultaneously. Therefore, their contribution averages out. This means that NSs with such a surface thermal distribution cannot produce a pulsed fraction above $\approx$10\%. They could, however, contain significant internal magnetic energy. Thus, these NSs could emit more in thermal surface radiation than it is normally expected from NS of a similar age. Moreover, since the spin-down is caused by the dipolar component only, the~field estimates based on measurements of the period and its derivative can be orders of magnitude lower than the actual maximum field value. Therefore such NSs with small-scale magnetic field provide a viable explanation for properties of central compact objects \citep{2020MNRAS.495.1692G,Igoshev2021ApJ}.

\section{Battery~Effects}\label{sec:battery}
As discussed in the previous section, the~large conductivity in a NS crust makes so that its thermal structure and evolution tends to follow that of the magnetic field. This allows one to use the thermal structure of the crust as a direct and, to~a certain extent, observable realisation of the Hall evolution. However, temperature itself affects the induction equation both through the thermal dependence of microphysical quantities (electric conductivity, electron relaxation time, etc.) and thermoelectric effects~proper.

The latter appear in the presence of a temperature gradient, and~are in general quite weak. The~application of these mechanisms to compact object astrophysics dates back to \citet{1980SvA....24..177D} (for white dwarfs) and \citet{1983MNRAS.204.1025B} (who extended the analysis to NSs). Thermoelectric effects were studied with the Hall effect in great mathematical detail in a series of papers by Geppert and Wiebicke~\cite{1991A&AS...87..217G,1991A&A...245..331W,1992A&A...262..125W,1995A&A...294..303W,1995A&A...300..429G,1996A&A...309..203W}, which laid the foundations for many magnetothermal evolution models (e.g., \citep{2007A&A...470..303P}, see also  \citet{2019LRCA....5....3P} for a review), but~they ended up being quite overlooked in the last decades. However, the~necessity of revisiting them more systematically  has recently become apparent for at least two reasons: first, computational advancements \citep{Igoshev2021NatAs} showed that the thermal structure of a NS crust can be quite variegated; second, transient bursting phenomena are often linked to some sort of heat deposition in the crust other than the ohmic dissipation that locally enhance thermal gradients \citep{2012ApJ...P}. Although~the exact mechanisms underlying these events are not fully understood, the~strong temperature gradients they produce may be the source of the additional field. This is of particular interest in the context of highly magnetised environments, i.e.,~in the modelling of the transient activity of magnetars \citep{2015RPPh...78k6901T}.

The physics underlying thermoelectric effects is again encapsulated by the basic fact that electrons are the main carriers of both electric currents and thermal energy in a NS crust. Thus, a~temperature gradient alters the charge balance, generating an electric field given by (e.g., \citep{ziman1972principles})
\begin{equation}
    \vec{E}=\hat{G}\cdot\vec{\nabla}T
\end{equation}
where the tensor $G$ is the so called thermopower.  
 In~general, it is composed of an isotropic part (the Seebeck term) and of anisotropic one (the Ettingshausen-Nernst term); the simplest choice is to only consider the former, which for a completely degenerate Fermi gas gives $G_{ij}=-S_e/e\, \delta_{ij}$; this form has been used to write the last term in Equation~(\ref{eq:heat}). This electric field enters the induction Equation~(\ref{HALL_EQ}) as an additional term, so that it is able to generate a magnetic field; this process is known as the Biermann battery effect \citep{1950ZNatA...5...65B}. The~Biermann battery is able to amplify to a great extent a small seed field, and~as such is a popular mechanism to explain magnetic structure formation on cosmological scale (e.g., \citep{2000ApJ...539..505G}) and in relativistic jets (e.g., \citep{2019A&A...622A.122P}). 

\begin{figure}[H]
\subfigure[]{\includegraphics[width=.41\linewidth]{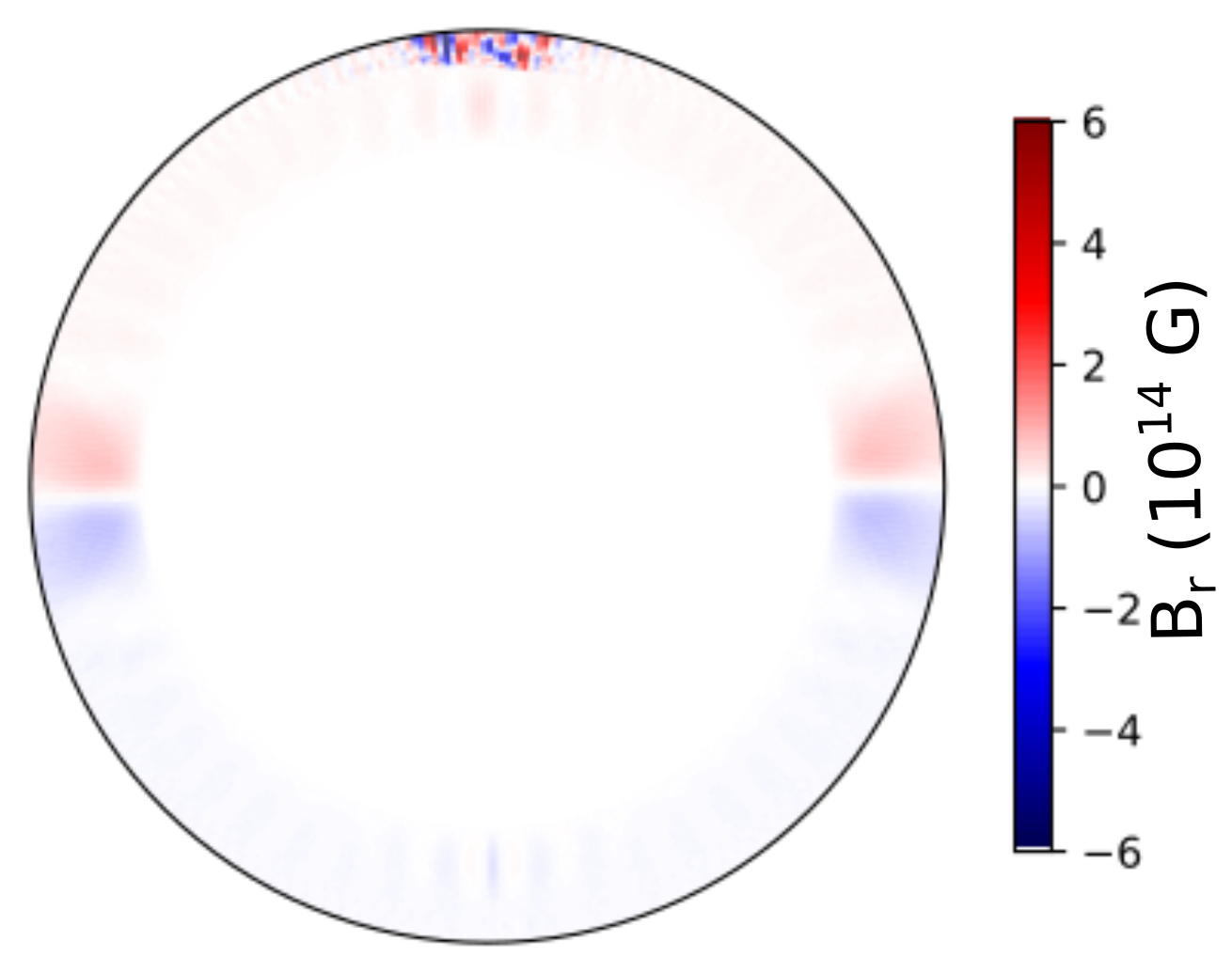}}~~
\subfigure[]{\includegraphics[width=.41\linewidth]{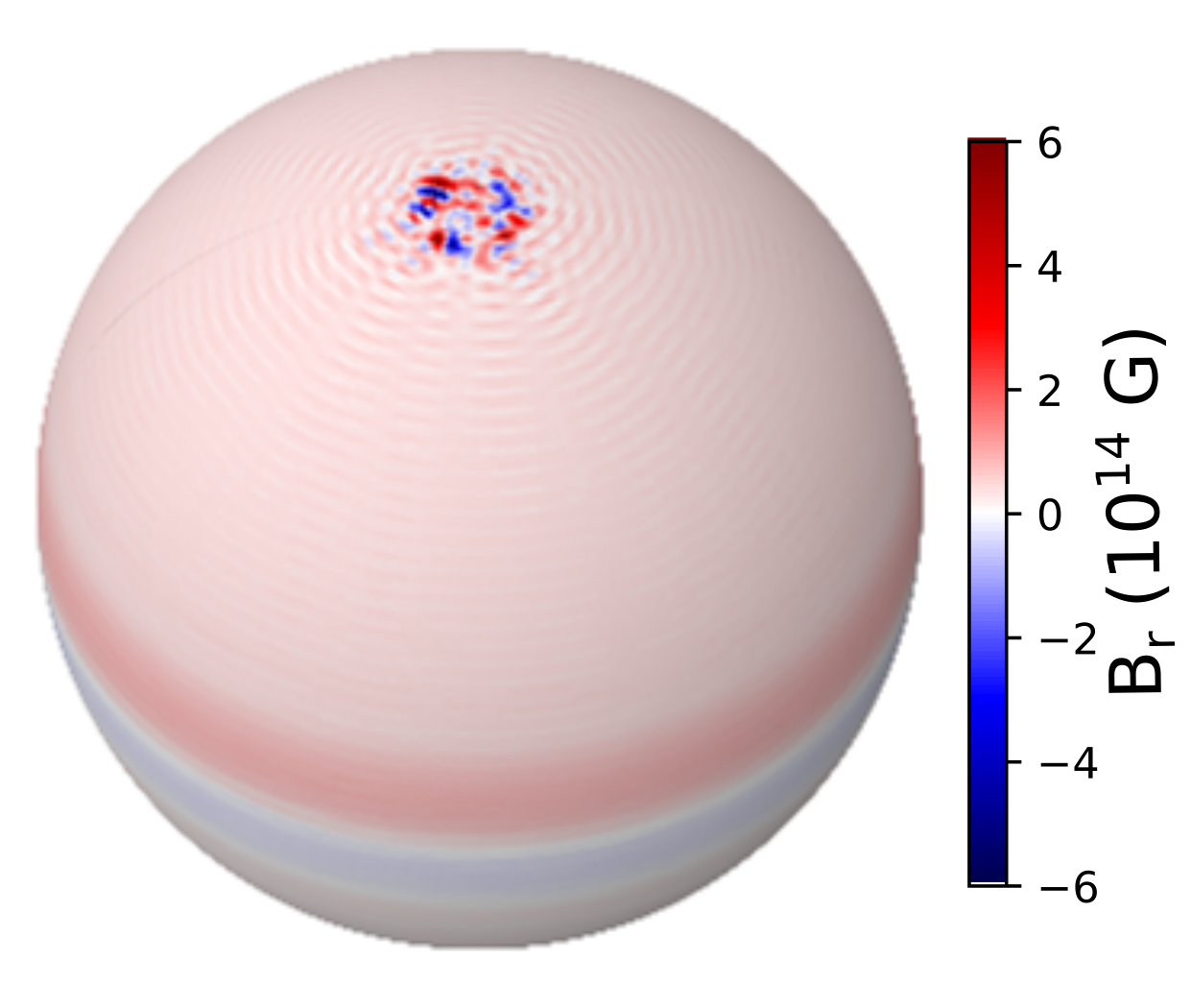}}\\~~

    \subfigure[]{\includegraphics[width=.79\linewidth]{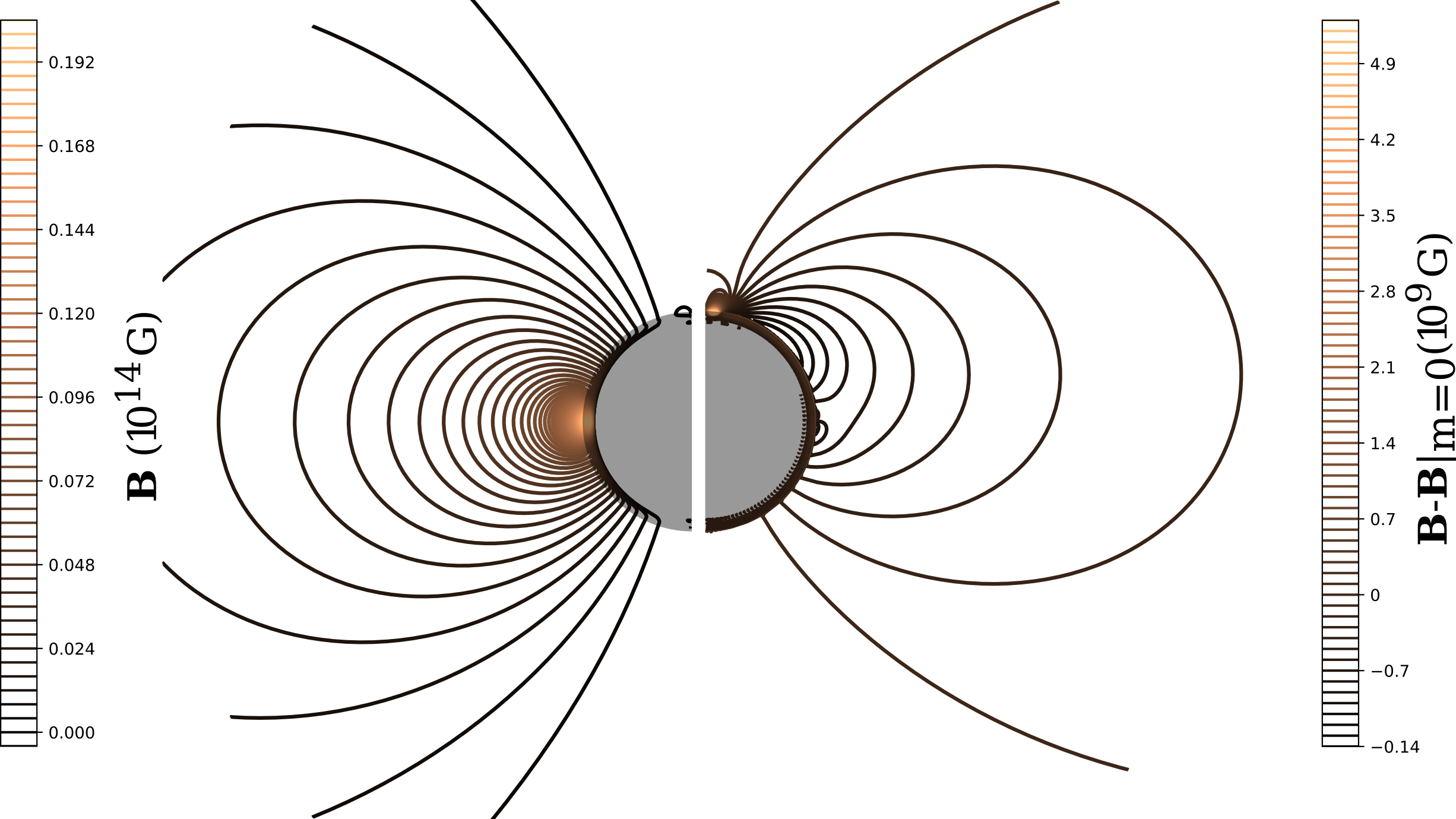}}
    \caption{Effect of the battery term after a episode of heat injection in a small patch of the crust. Panel (a): the radial component of the magnetic field $B_r$ shown in a cut along the prime meridian (crust width $4\times$ for visualisation); Panel (b): $B_r$ shown on a projection of the surface; Panel (c): magnetic field lines in the magnetosphere obtained by continuation of the boundary condition $\nabla\times B(R)=0$. The right half shows the lines of the asymmetric component on the prime meridian, obtained by subtracting the $m=0$ component which is dominated by the dipolar field. Figure adapted from~\cite{2020ApJ...903...40D}, Copyright AAS. Reproduced with permission.}
    \label{f:battery}
\end{figure}

A realisation of an efficient battery effect in the crust of a NS can be expected in cases when strong temperature gradients are established by a mechanism not contained in the usual magnetothermal evolution equations. The~energy source could still be the magnetic field itself, if~it undergoes phenomena of reconnection, or~come from external agents like accretion or backflowing magnetospheric currents. 
As a model of the latter case, Figure~\ref{f:battery} shows the effect of heating a circular patch of radius $\sim$1\,km in the outermost region of the crust with $H=5\times10^{26}\,$erg/s for a few years, until~a quasi stationary state is reached. The~initial state was that of a purely dipolar field, which had previously been evolved \linebreak for $\sim$10$^5$\,yr (see~\cite{2020ApJ...903...40D} for more details). The~additional temperature gradient builds up a sizeable magnetic field, which can be extrapolated to the low magnetosphere to give rise to small scale magnetic loops. Indeed, small scale magnetic loops have been invoked to explain the observed absorption features in some magnetars~\cite{2013Natur.500..312T}, even though a comprehensive analysis of these phenomena in terms of battery effects has not been attempted as yet due to the numerical hindrances associated to it (see e.g., \cite{2015ApJ...802...43G}). 

\vspace{-14pt}

\section{Conclusions}
\label{sec:conclusions}

When the crucial role of the magnetic field in the physics of NSs was assessed, it was modelled as a dipolar field, unchanged for the entire life of the star. It was soon realised that the role of Ohmic decay on astrophysically relevant timescales was crucial, establishing the idea that the field may  indeed evolve. Models then turned to the Hall effect that, while not dissipating the field, is able to alter it and enhance Ohmic decay. 
The study of the impact of the Hall effect in the magnetic field evolution of NSs has been a successful endeavour. It has resolved many questions related to the nature of NSs, observable quantities, and~it has also motivated studies in the areas of Applied Mathematics, fluid dynamics and computational physics. Yet, some of the physical assumptions postulated for the development of Hall-Ohmic equations may not hold in their entirety, especially in the most extreme cases, either in the NSs with the strongest magnetic fields, the~hottest ones, or~the ones with sharp temperature~gradients. 

In this review, we have focused on effects that are beyond the standard Hall evolution in the crust. Namely, we have studied the impact of the effects of a plastic flow once the magnetic field is substantially strong so that the crust fails, the~magneto-thermal evolution, where the interplay of the heat released by the Ohmic decay and its transport mediated by the magnetic field are discussed and battery effects, arising from temperature gradients. We have shown that these phenomena widely enrich the evolutionary paths of the magnetic field and their observable appearance. We remark that, while great progress has been made so far, there are still areas for progress. Plastic flows have been studied only in axisymmetric and plane-parallel cartesian geometries, and~a full 3D model is yet to be resolved. The~magnetothermal evolution has been implemented in 3D studies, nevertheless, the~huge parameter space of microphysical properties, such as variations of the thermal and electric conductivities, is still to be explored. Finally, battery effects have been studied in a small subset of cases and they clearly deserve a more detailed exploration of the many different mechanisms that can generate temperature gradients. We strongly believe that such studies will widen the horizons and pave the way for a clear comprehension of NSs, both in terms of their phenomenology and of their use as benchmarks for fundamental physics~applications.

%%%%%%%%%%%%%%%%%%%%%%%%%%%%%%%%%%%%%%%%%%

\vspace{6pt} 

%%%%%%%%%%%%%%%%%%%%%%%%%%%%%%%%%%%%%%%%%%
%% optional
%\supplementary{The following are available online at \linksupplementary{s1}, Figure S1: title, Table S1: title, Video S1: title.}


%%%%%%%%%%%%%%%%%%%%%%%%%%%%%%%%%%%%%%%%%%
\authorcontributions{Conceptualization, K.N.G., D.D.G. and A.I.; writing---original draft preparation, K.N.G., D.D.G. and A.I.; writing---review and editing, K.N.G., D.D.G. and A.I. All authors have read and agreed to the published version of the~manuscript.}

\funding{K.N.G. acknowledges funding from grant FK 81641 ``Theoretical and Computational Astrophysics'', ELKE. A.I. acknowledges funding from STFC grant no. ST/S000275/1.} %AUTHORS: We have added funding information for A.I.

\institutionalreview{Not applicable. } %MDPI: Please add sections below. 
%{In this section, please add the Institutional Review Board Statement and approval number for studies involving humans or animals. Please note that the Editorial Office might ask you for further information. Please add ``The study was conducted according to the guidelines of the Declaration of Helsinki, and approved by the Institutional Review Board (or Ethics Committee) of NAME OF INSTITUTE (protocol code XXX and date of approval).'' OR ``Ethical review and approval were waived for this study, due to REASON (please provide a detailed justification).'' OR ``Not applicable'' for studies not involving humans or animals. You might also choose to exclude this statement if the study did not involve humans or animals.}

\informedconsent{Not applicable.}
%{Any research article describing a study involving humans should contain this statement. Please add ``Informed consent was obtained from all subjects involved in the study.'' OR ``Patient consent was waived due to REASON (please provide a detailed justification).'' OR ``Not applicable'' for studies not involving humans. You might also choose to exclude this statement if the study did not involve humans.

%Written informed consent for publication must be obtained from participating patients who can be identified (including by the patients themselves). Please state ``Written informed consent has been obtained from the patient(s) to publish this paper'' if applicable.}

\dataavailability{Numerical data related to this study are available upon request.} 

%\acknowledgments{Thanks}

\conflictsofinterest{The authors declare no conflict of~interest.}

\begin{adjustwidth}{-\extralength}{0cm}

\reftitle{References}

\end{adjustwidth}

\end{document}